\documentclass[aps,prx,twocolumn,showpacs]{revtex4-1}
\usepackage{bm}
\usepackage{mathrsfs}
\usepackage{amsmath}
\usepackage{amssymb}
\usepackage{graphicx}
\usepackage{amsfonts}
\usepackage{amsthm}
\usepackage{color}
\usepackage{dcolumn}
\usepackage{txfonts}
\usepackage[caption=false]{subfig}
\usepackage[T1]{fontenc}

\DeclareMathOperator{\cov}{cov}

\DeclareMathOperator\erf{erf}

\begin{document}

\title{Sub-microsecond high-fidelity dispersive readout of a spin qubit with squeezed photons}
\author{Chon-Fai Kam}
\author{Xuedong Hu}
\email{Corresponding author. Email: xhu@buffalo.edu}
\affiliation{Department of Physics, University at Buffalo, SUNY, Buffalo, New York 14260, USA}

\begin{abstract}
Fast and high-fidelity qubit measurement is essential for realizing quantum error correction, which is in turn a key ingredient to universal quantum computing.  For electron spin qubits, fast readout is one of the significant road blocks toward error correction.  Here we examine the dispersive readout of a single spin in a semiconductor double quantum dot coupled to a microwave resonator. We show that using displaced squeezed vacuum states for the probing photons can improve the qubit readout fidelity and speed.  Under condition of proper phase matching, we find that a moderate, and only moderate, squeezing can enhance both the signal-to-noise ratio and the fidelity of the qubit-state readout, and the optimal readout time can be shortened to the sub-microsecond range with above $99\%$ fidelity.  These enhancements are achieved at low probing microwave intensity, ensuring non-demolition qubit measurement.
\end{abstract}

\maketitle
\textit{Introduction}.--- Significant progress has been made toward building a universal quantum computer over the past decade, based on a variety of qubit platforms \cite{kjaergaard2020superconducting, chen2021exponential, bruzewicz2019trapped, wright2019benchmarking, burkard2021semiconductor, noiri2022fast, xue2022quantum, mkadzik2022precision, takeda2022quantum, takeda2019toward, bourassa2021blueprint, madsen2022quantum}. Among these options, spin qubits in silicon-based semiconductor quantum dots, while lagging somewhat behind other systems such as superconducting and trapped ion qubits \cite{egan2021fault, riel2022quantum}, hold significant long-term potential due to their excellent quantum coherence \cite{burkard2021semiconductor}, a small qubit footprint, and compatibility with the well-established integration techniques of the microelectronics industry \cite{gonzalez2021scaling}. Further development of spin qubits still faces a multitude of challenges.  For example, a key element of scalable quantum computing is the implementation of active quantum error correction (QEC) \cite{cramer2016repeated}, which requires repeated measurements of physical qubits \cite{terhal2015quantum} and real-time feedback \cite{saraiva2022dawn}. The success of active QEC hinges on high-fidelity readout in times significantly shorter than the qubits' decoherence times \cite{gonzalez2021scaling}.  For a spin qubit in isotopically purified silicon, spin dephasing time is typically in the order of tens of microseconds \cite{burkard2021semiconductor}, making it crucial to attain readout fidelity above the $99\%$ threshold of the surface code for QEC \cite{raussendorf2007fault, fowler2012surface} in a sub-microsecond time scale. 

Commonly used spin measurement approaches based on spin-dependent tunneling (the so-called Elzerman technique) \cite{elzerman2004single} or spin blockade \cite{petta2005coherent}, together with a DC charge sensor, are too slow for active QEC.  A Radio Frequency (RF) version of the charge sensor, such as the RF single-electron transistor (RF-SET), could speed up the measurement, achieving single-shot readout of a single-spin qubit with 97\% fidelity in 1.5 $\mu$s \cite{keith2019single} and 99.9\% fidelity in 6 $\mu$s \cite{curry2019single}.  Recently, a single-shot singlet-triplet readout based on RF-reflectometry has achieved a signal-to-noise ratio (SNR) of 6 in 0.8 $\mu$s \cite{noiri2020radio}, while another reached 99\% fidelity in 1.6 $\mu$s \cite{connors2020rapid}. However, the inclusion of charge sensors does increase the complexity of a device, and the larger footprint of RF-SETs constrains their placement in highly connected qubit architectures \cite{de2023silicon}.  Gate-reflectometry-based dispersive spin readout skip the charge sensor and send the RF probing pulse directly to the qubit through one of its gates \cite{hu2019fast}. Using an off-chip resonator, readout fidelities ranging from 73.3\% to 98\% have been achieved, though achieving single-shot readout has so far required integration times on the order of milliseconds \cite{pakkiam2018single, west2019gate, urdampilleta2019gate}. On the other hand, with an on-chip resonator and a microwave frequency probe pulse, a single-shot readout fidelity of 98\% has been achieved at a respectable $6\:\mu$s measurement time \cite{zheng2019rapid}. In short, fast readout with high fidelity and small footprint remains a formidable road block for scalable quantum information processing based on spin qubits.

One possible approach to improve a reflectometry-based spin readout protocol is to employ squeezed states for photons in the measurement process rather than the conventional coherent states. Squeezed states are nonclassical states known for modified quantum noise profiles \cite{walls1983squeezed, zhang1990coherent, kam2023coherent}. By introducing quantum correlations among photons, quantum fluctuations are periodically reduced to below the standard quantum limit in one field quadrature component while simultaneously increased in the other quadrature \cite{andersen201630}. Squeezed states have been extensively studied in various research fields over the last decade. For instance, high intensity squeezed light has been employed in the latest laser-based gravitational wave detectors \cite{tse2019quantum, acernese2019increasing}, resulting in a nearly ten-fold increase in sensitivity \cite{dwyer2022squeezing}.  In the field of quantum information processing, squeezed state has been applied in continuous-variable quantum key distribution \cite{gehring2015implementation}, quantum sensing \cite{lawrie2019quantum}, and high-precision cavity spectroscopy \cite{junker2021high}.  Squeezed states have helped enhance the signal-to-noise ratio (SNR) of superconducting qubit readout \cite{barzanjeh2014dispersive}, leading to a 24\% increase in the final SNR for superconducting qubit measurement \cite{eddins2018stroboscopic}.  Furthermore, a 31\% enhancement in the SNR for superconducting qubit readout with 99\% fidelity was realized when a two-mode squeezed vacuum is used \cite{liu2022noise}. 

Here we explore the impact of squeezing on the dispersive readout of a spin qubit coupled to an on-chip microwave resonator, and its subsequent back-action effects on the spin qubit. We find that using a low-intensity (so as to reduce decohering effect of the photons on the qubit) displaced squeezed vacuum state could yield significant enhancements in the SNR, enabling rapid and high-fidelity dispersive readout of the spin qubit through standard techniques.  Interestingly, we also find that only modest degree of squeezing, under proper phase-matching conditions, improves spin measurement. Larger squeezing actually leads to a deterioration of measurement SNR and fidelity due to contributions from the ``anti-squeezed'' quadrature.

\begin{figure}[tbp]
\includegraphics[width=0.45\columnwidth]{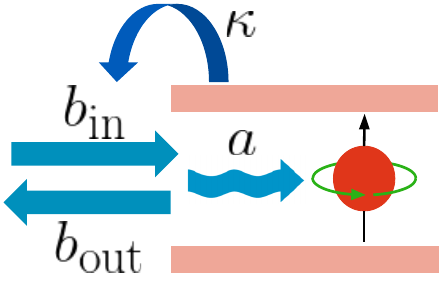}
\caption{Schematic of the setup for dispersive detection of spin qubits.}
\label{S0}
\end{figure}

\textit{The model}.--- We consider dispersive readout of a spin qubit assisted by a single mode microwave resonator. In the rotating frame, the dispersive coupling Hamiltonian between the qubit and the resonator field is \cite{d2019optimal}
\begin{equation}\label{Dispersive}
    H_s = \frac{1}{2}(\delta_s-\chi_s)\sigma_z+\left(\delta_c-\chi_s\sigma_z\right)a^\dagger a,
\end{equation}
where $\delta_c$ and $\delta_s$ denote the detunings of the probe from the resonator and the spin qubit, respectively, and $\chi_s$ is the dispersive coupling strength between the spin qubit and the resonator. The photons can enter and leave the resonator through a single input/output port with a leakage rate $\kappa$, governed by an interaction of the form \cite{d2019optimal}:
\begin{equation}\label{Interaction}
V_{\mathrm{in}}=i\sqrt{\kappa}(b_{\mathrm{in}}^\dagger a-b_{\mathrm{in}}a^\dagger), 
\end{equation}
where $b_{\mathrm{in}}$ and $b_{\mathrm{in}}^\dagger$ are the annihilation and creation operators of the input radiation field mode. While multiple input radiations may be possible through multiple ports, we focus on the case of a single input/output port for simplicity. From the effective Hamiltonian $H_{\mathrm{eff}} \equiv H_s+V_{\mathrm{in}}$ for the coupled qubit-photon system, one can derive the Langevin equation of motion for the resonator field as
\begin{equation}\label{Master}
    \dot{a} = -\left[i\left(\delta_c-\chi_s\sigma_z\right)+\frac{\kappa}{2}\right]a  - \sqrt{\kappa}b_{\mathrm{in}},
  \end{equation}
where the first term on the right-hand side describes the dispersive shift of the resonator field as well as damping, while the last term represents the driving of the resonator through its input port. The damping term in Eq.\:\eqref{Master} is a result of the Markovian approximation, depending solely on the system operators at the current time. It forms the basis of the input-output theory \cite{gardiner1985input, gardiner2004quantum}. Under the Markovian approximation, the output radiation field mode can be determined by the input-output relation \cite{gardiner1985input}: 
\begin{equation} \label{IORelation}
b_{\mathrm{out}} = b_{\mathrm{in}} + \sqrt{\kappa} a,
\end{equation} 
where $b_{\mathrm{out}}$ and $b_{\mathrm{out}}^\dagger$ are the annihilation and creation operators of the output radiation field mode respectively. To ensure the commutation relation $[b_{\mathrm{out}},b_{\mathrm{out}}^\dagger]=[b_{\mathrm{in}},b_{\mathrm{in}}^\dagger]$, $[b_{\mathrm{in}}^\dagger,a] = [a^\dagger, b_{\mathrm{in}}] = \frac{\sqrt{\kappa}}{2}[a, a^\dagger]$ are required \cite{gardiner1985input}. 

The Langevin equation Eq.\:\eqref{Master} and the input-output relation Eq.\:\eqref{IORelation} form the basis of our study. The former describes how the resonator field is affected by the qubit and the pumping from the outside, while the latter relates the reflected field to the input and the resonator field, allowing us to evaluate the effect of the qubit state on the output. When the qubit-resonator interaction is dispersive, the spin measurement becomes non-destructive when the resonator field is weak, wherein $\sigma_z(t)$ is approximately a constant of motion, i.e., $\sigma_z(t)\approx \sigma_z(0)$. Consequently, $\sigma_z$ can be represented as a real number $\sigma=\pm 1$ onwards, allowing for an analysis of the readout contrast. Without loss of generality, we assume $\delta_c=0$ from now on.

\begin{figure}[tbp]
	\subfloat{\includegraphics[width=0.5\columnwidth]{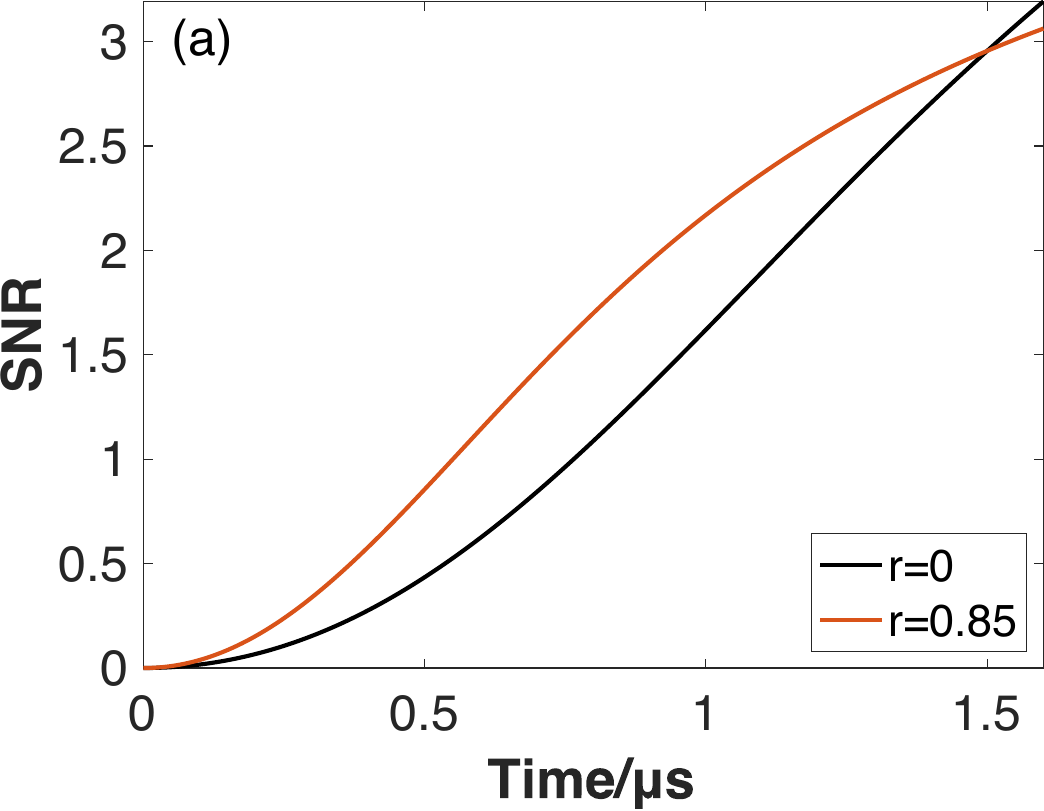}}\label{sfig:S1a}\hfill
	\subfloat{\includegraphics[width=0.5\columnwidth]{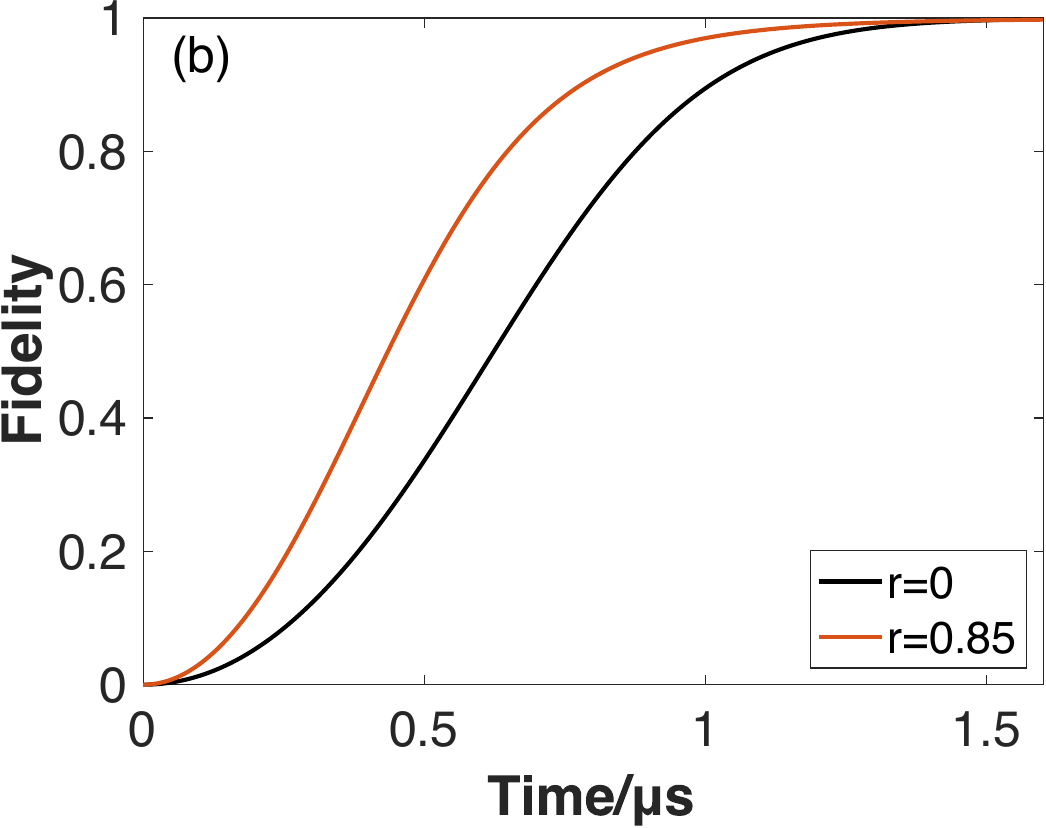}}\label{sfig:S1b}\hfill
	\subfloat{\includegraphics[width=0.5\columnwidth]{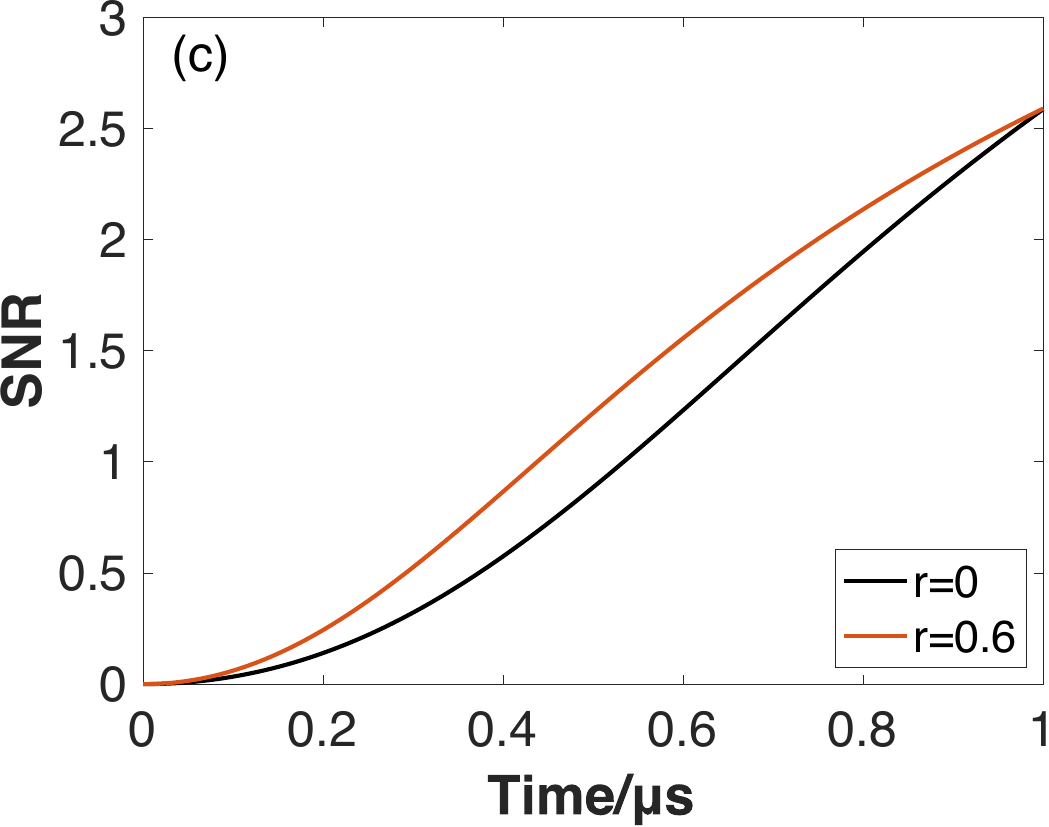}}\label{sfig:S1c}\hfill
	\subfloat{\includegraphics[width=0.5\columnwidth]{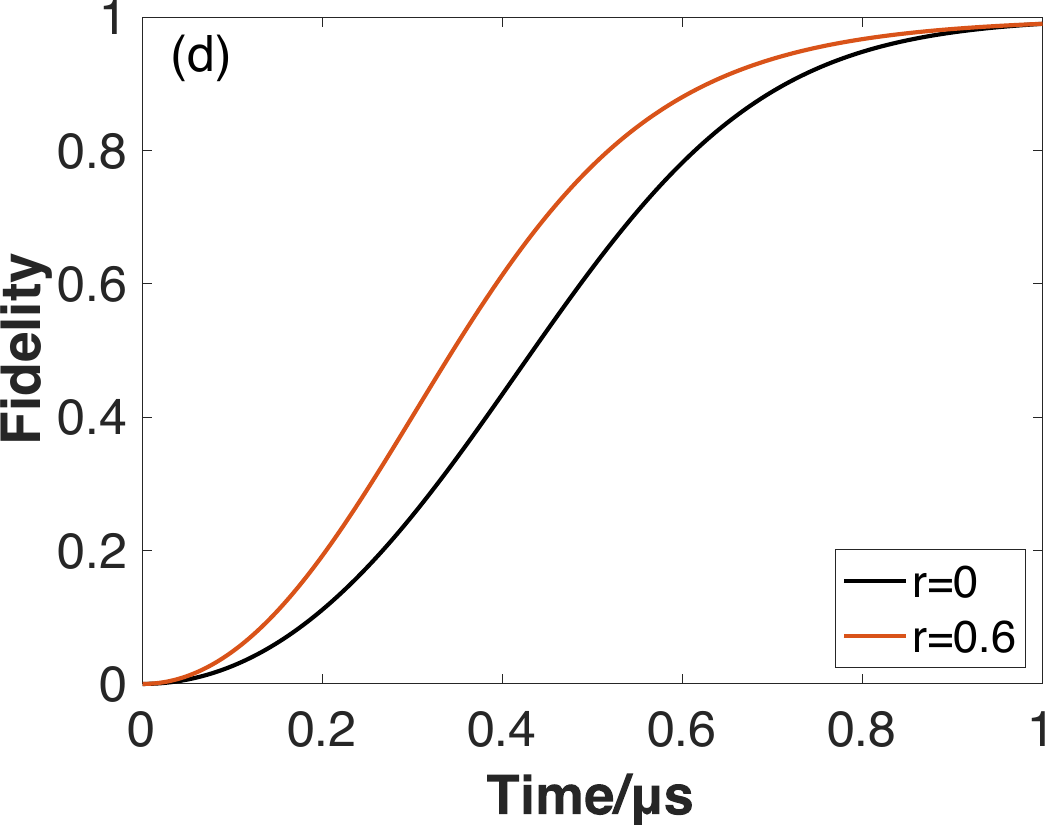}}\label{sfig:S1d}
\caption{SNR and readout fidelity with respect to different squeezing parameters and leakage rates. Here, $\alpha=\sqrt{30}$, $\chi_s=2\pi\times 0.15\:\mbox{MHz}$, $\theta_\xi=\pi$, and $T_1=3\:\mbox{ms}$. The leakage rates for the panels (a), (b) and (c), (d) are $\kappa=\chi_s$ and $\kappa=2\chi_s$ respectively. The black lines represent the coherent state inputs.}
\label{fig:S1}
\end{figure}

In RF-reflectometry, the output field $b_{\mathrm{out}}$ is sent through a phase-preserving amplifier \cite{clerk2010introduction}, which detects both quadratures equally well, and is subsequently measured using a homodyne detector \cite{yuen1983noise} by mixing with a local oscillator with phase $\varphi$. The resulting photocurrent $I\propto \frac{1}{\sqrt{2}}(b_{\mathrm{out}}e^{-i\varphi}+b_{\mathrm{out}}^\dagger e^{i\varphi})\equiv Q_{\mathrm{out}}\cos\varphi+P_{\mathrm{out}}\sin\varphi$.  By choosing the local oscillator phase to be $\varphi=0$ or $\pi/2$, the photocurrent would be proportional to $Q_{\mathrm{out}}$ or $P_{\mathrm{out}}$, respectively. We thus rewrite the resonator Langevin equation for the resonator field in terms of their quadratures as
\begin{equation}\label{LangevinQ}
\dot{X}(t)=
M X(t)
-\sqrt{\kappa}X_{\mathrm{in}},
M\equiv -\frac{\kappa}{2}I\mp i \chi_s \tau_y,
\end{equation}
where $X\equiv(Q,P)^\top$, $X_{\mathrm{in}}\equiv(Q_{\mathrm{in}},P_{\mathrm{in}})^\top$, and $\tau_y$ is the Pauli matrix in the $y$-direction acting on the field quadratures. Here, $Q$ and $P$ are referred to as the amplitude and phase quadratures of the resonator field, and $Q_{\mathrm{in}}$ and $P_{\mathrm{in}}$ are those of the input radiation field. For a continuous wave input radiation field, Eq.\:\eqref{LangevinQ} is solved as
\begin{gather}
X(t) = e^{Mt}X(0)-\sqrt{\kappa}\int_0^tds e^{Ms} X_{\mathrm{in}},
\end{gather}
where $e^{Mt}=f(t)I\mp ig(t)\tau_y$, $f(t)\equiv e^{-\frac{\kappa}{2}t}\cos \chi_s t$, and $g(t)\equiv e^{-\frac{\kappa}{2}t}\sin\chi_s t$.  Substituting this result into Eq.\:\eqref{IORelation} would then yield the solutions for the output field quadratures.

\textit{Signal-to-noise ratio}.--- Recall that in RF reflectometry, the homodyne measurement yields a photocurrent proportional to the expectation value of an output field quadrature. Without loss of generality, here we choose the output quadrature as $P_{\mathrm{out}}\equiv P_{\mathrm{in}}+\sqrt{\kappa}P$, which is specified by a local oscillator phase of $\pi/2$. To quantify the distinguishability between the two qubit states, we introduce the signal-to-noise ratio (SNR) based on the measurement outcomes. It is defined as the ratio of the contrast between output signals from the two qubit states to the sum of the associated standard deviations:
\begin{equation}\label{SNRDef}
\mbox{SNR}(t)\equiv\frac{|\langle \mathscr{M}^{+1}(t)\rangle-\langle \mathscr{M}^{-1}(t)\rangle|}{\Delta \mathscr{M}^{+1}(t)+\Delta \mathscr{M}^{-1}(t)}.
\end{equation}
Here, $\langle \mathscr{M}^{\pm}(t)\rangle$ and $\Delta \mathscr{M}^{\pm 1}(t)$ are the expectation values and standard derivations, respectively, of the time-integrated output quadratures $\mathscr{M}^{\pm 1}(t)\equiv \int_0^t P^{\pm 1}_{\mathrm{out}}(s)ds$, where the superscript is used to denote the two qubit states. Particularly, the time-integrated output quadratures takes the explicit form of
\begin{equation}\label{Signal}
    \mathscr{M}^{\pm}(t) =A(t) P_{\mathrm{in}} \mp B(t )Q_{\mathrm{in}}+\sqrt{\kappa}F(t)P(0)\pm \sqrt{\kappa}G(t)Q(0),
\end{equation}
where the time-dependent coefficients are $A(t)\equiv t-\kappa\int_0^t F(s)ds$, $B(t)\equiv \kappa \int_0^t G(s)ds$, $F(t)\equiv\int_0^t f(s)ds$, and $G(t)\equiv\int_0^t g(s)ds$. According to Eq.\:\eqref{Signal}, when the resonator field is initially in a vacuum state, the contrast between the output signals is given by $2|B(t)\langle Q_{in}\rangle|$, which depends solely on the input quadrature $Q_{\mathrm{in}}$, and is independent of $P_{\mathrm{in}}$. 

Now we are ready to explore the effect of squeezing in the input field.  Specifically, we choose a displaced squeezed vacuum state $|\boldsymbol{\alpha},\boldsymbol{\xi}\rangle \equiv D(\boldsymbol{\alpha})S(\boldsymbol{\xi})|0\rangle$ as an input, where $D(\boldsymbol{\alpha})\equiv\exp\int dk (\alpha_k b_k^\dagger-\alpha_k^*b_k)$ is a continuous displacement operator, and $S(\boldsymbol{\xi}) \equiv \exp\frac{1}{2}\int dk(\xi_k^* b_k^2-\xi_k b_k^{\dagger 2})$ is a continuous squeezing operator \cite{fedorov2016displacement, mandel1995optical}. Here $\boldsymbol{\alpha} = \alpha e^{i\theta_\alpha}$ and $\boldsymbol{\xi} = \xi e^{i\theta_\xi}$. In this state, the expectation values of the annihilation and creation operators coincide with those of coherent states: $\langle b_{\mathrm{in}}\rangle=\alpha(t)$ and $\langle b^\dagger_{\mathrm{in}}\rangle=\alpha^*(t)$ (See Supplemental Material). As an alternative, one can also use the squeezed coherent state $|\boldsymbol{\xi},\boldsymbol{\gamma}\rangle\equiv S(\boldsymbol{\xi})D(\boldsymbol{\gamma})|0\rangle$ as an input field. It can be viewed as a displaced squeezed vacuum state with the same squeezing parameters, but featuring distinct displacement amplitudes $\alpha_k=\gamma_k\cosh r_k-\gamma_k^*\sinh r_k e^{i\theta_{\xi_k}}$.

\begin{figure}[tbp]
	\subfloat{\includegraphics[width=0.495\columnwidth]{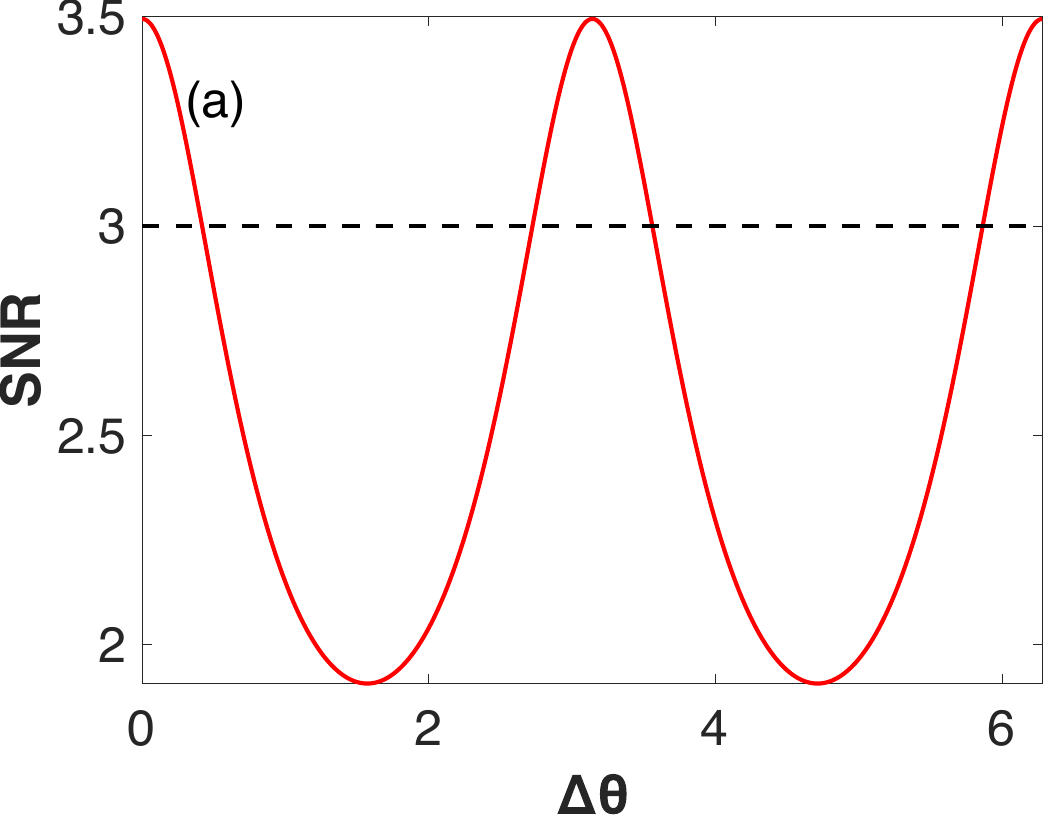}}\label{sfig:S4a}\hfill
	\subfloat{\includegraphics[width=0.495\columnwidth]{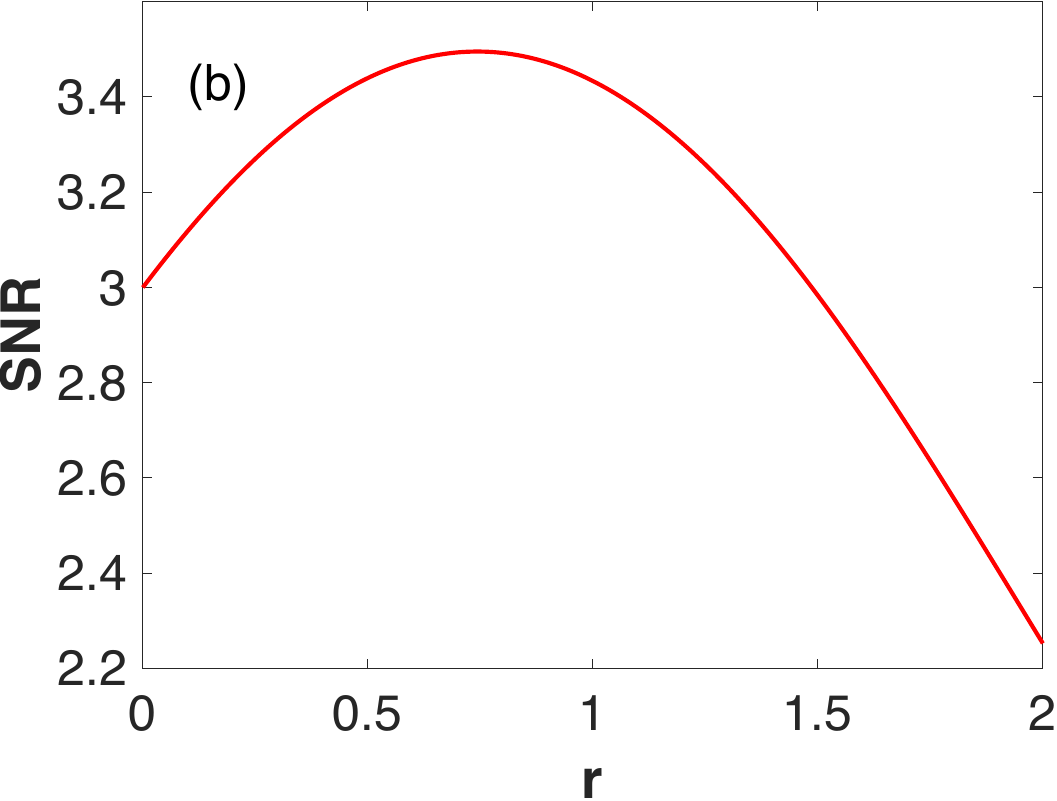}}\label{sfig:S4b}\hfill
	\subfloat{\includegraphics[width=0.495\columnwidth]{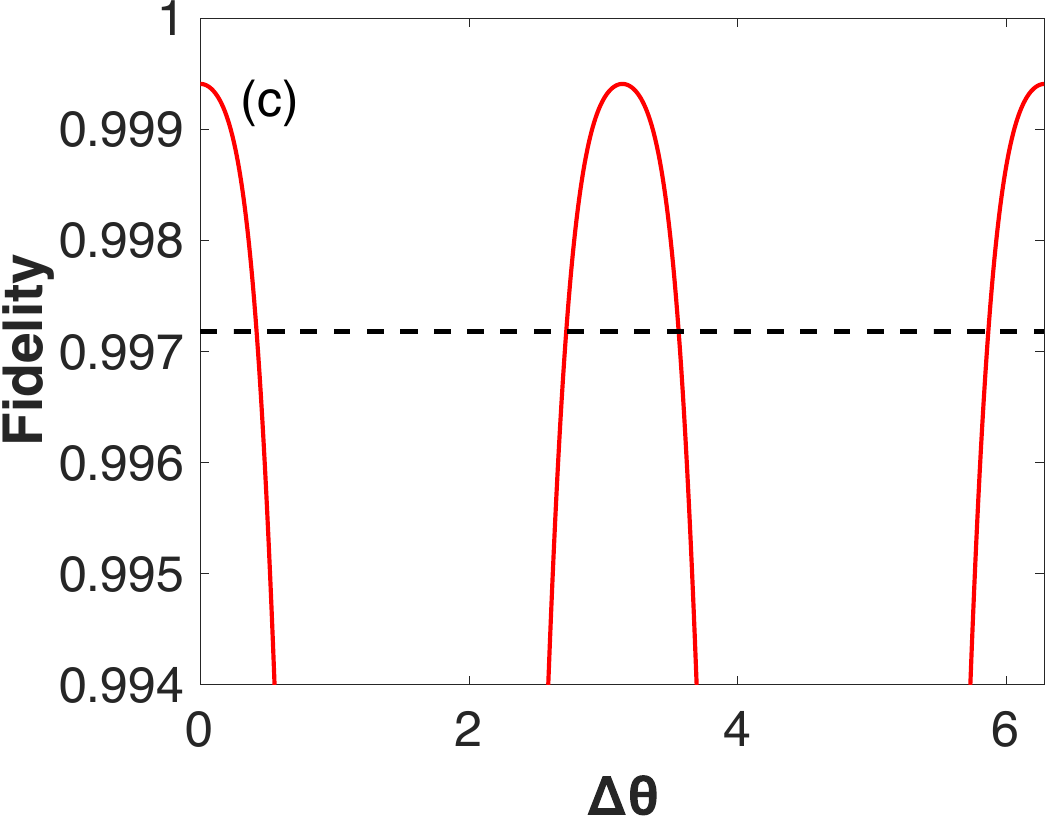}}\label{sfig:S4c}\hfill
	\subfloat{\includegraphics[width=0.495\columnwidth]{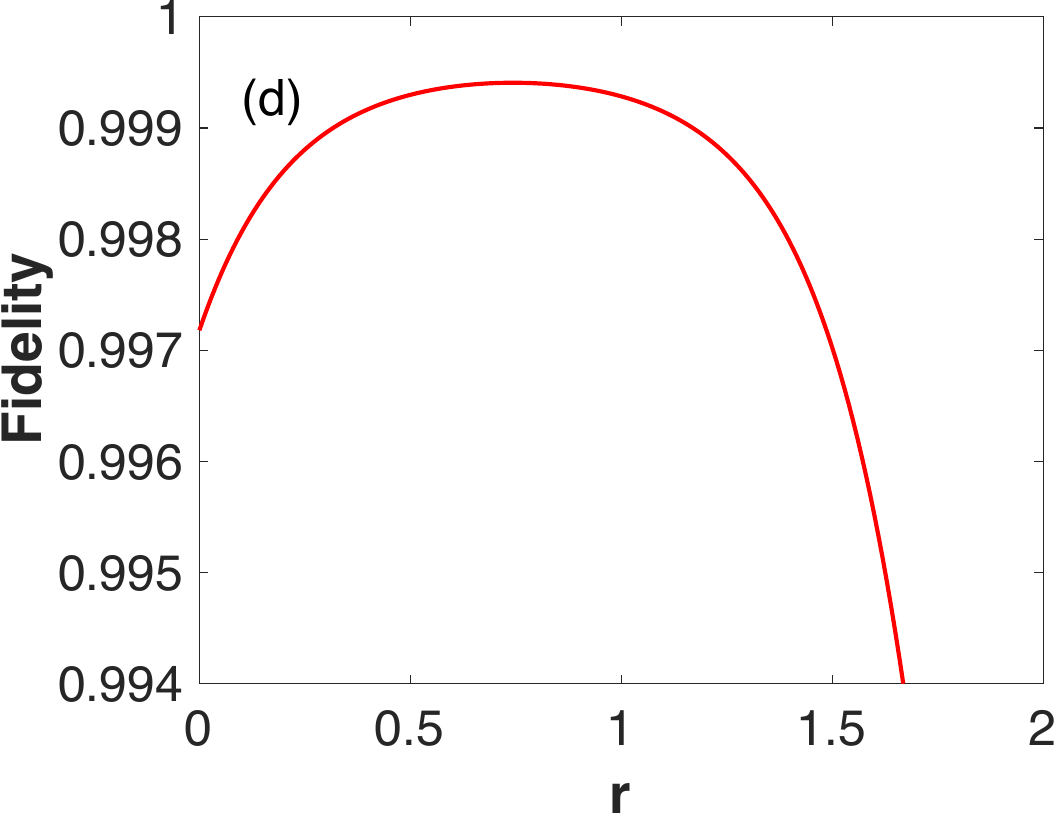}}\label{sfig:S4d}
\caption{SNR and readout fidelity with respect to different phase mismatches and squeezing parameters at a fixed measurement time $t\approx 0.714\:\mu s$. Here, $|\alpha|=10$, $\kappa=2\chi_s$, $\chi_s=2\pi\times 0.15\:\mbox{MHz}$, and $T_1=3\:\mbox{ms}$. For Panels (a) and (c), a squeezing parameter of $r = 0.74$ is chosen, and the black dash lines represent the coherent state inputs. For Panels (b) and (d), the phase matching condition is given by $\Delta \theta \equiv \phi - \theta_\xi /2 = 0$.}
\label{fig:S4}
\end{figure}

Consider an example where the homodyne detection for quadrature $P_{\rm out}$ with a local oscillator phase $\varphi=\pi/2$. In this case the signal contrast is proportional to $|\cos\theta_\alpha|$. To maximumize this contrast, $\theta_\alpha$ must vanish, which in turn allows us to obtain the standard deviations of the output quadrature for the two qubit states as
\begin{gather}
    (\Delta \mathscr{M}^{\pm}(t))^2=A^2(t)(\Delta P_{\mathrm{in}})^2+B^2(t)(\Delta Q_{\mathrm{in}})^2\nonumber\\
    +2A(t)B(t)\mbox{Re}(\cov(Q_{\mathrm{in}}P_{\mathrm{in}}))+\frac{\kappa}{2}(F^2(t)+G^2(t))\,.
\end{gather}
Given that the readout time is much shorter than $2/\kappa$, one can make the approximations $A(t)\approx t-\frac{1}{2}\kappa t^2$ and $B(t)\approx \frac{1}{6}\kappa\chi_s t^3$. Consequently, during a fast dispersive readout, the noise from the input quadrature $P_{\mathrm{in}}$ when leaving the resonator largely exceeds that from $Q_{\mathrm{in}}$. 
To enhance the SNR, it is thus beneficial to choose $P_{\mathrm{in}}$ as the squeezed quadrature, associated with $\theta_\xi=\pm\pi$, which leads to 
\begin{gather}\label{Noise}
    (\Delta \mathscr{M}^\pm(t))^2=\frac{e^{-2r}}{2}A^2(t)+\frac{e^{2r}}{2}B^2(t)+\frac{\kappa}{2}(F^2(t)+G^2(t)).
\end{gather}
For a fixed value of squeezing amplitude $r$, the optimal readout time is estimated to be $t\approx e^{-r}(6/\kappa\chi_s)^{1/2}$. 

In general, in a homodyne detection setup with the local oscillator phase $\varphi$ set to an arbitrary angle, the SNR of the chosen quadrature is still determined by Eq.\:\eqref{Noise}, but with $B(t)$ and $A(t)$ replaced by $B(t)\cos\Delta\theta+A(t)\sin\Delta\theta$ and $-B(t)\sin\Delta\theta+A(t)\cos\Delta\theta$, respectively. Here, $\Delta\theta \equiv \varphi-\theta_\xi/2$ is the phase mismatch between the local oscillator and the squeezing. In this general scenario, the signal contrast is proportional to $|\sin(\theta_\alpha-\varphi)|$. To maximize the contrast between the output signals, it is necessary for the phase of the displacement amplitude leads or lags that of the local oscillator by $\pi/2$. Besides, to minimize the signal noise from the input quadratures, it is beneficial to set the phase mismatch between the local oscillator and the squeezed state as an integer multiple of $\pi$. Hence, the optimal SNR enhancement is characterized by the phase matching relation
\begin{equation}\label{PhaseRelations}
\theta_\alpha-\varphi = \left(m+\frac{1}{2}\right) \pi \:\: \mbox{and}\:\: \varphi -\frac{\theta_\xi}{2}= m'\pi,
\end{equation}
where $m$ and $m'$ are arbitrary integers. Eq.\:\eqref{PhaseRelations} immediately leads to the more concise relation $2\theta_\alpha-\theta_\xi=(2(m+m^\prime)+1)\pi$, which gives the phase relationship required for squeezed input photon to help enhance the SNR for qubit readout.

\textit{Fidelity}.--- For an ideal qubit with an infinite (longitudinal) relaxation time $T_1$, the single-shot readout fidelity depends only on the SNR of the output signal. Specifically, when the standard derivations of the time-integrated output quadratures are equal ($\Delta \mathscr{M}^+(t)=\Delta \mathscr{M}^-(t)$), the single-shot readout fidelity $\mathscr{F}(t)$ is determined exclusively by the SNR as $\mathscr{F}(t) = \erf(\mbox{SNR}(t)/\sqrt{2})$, where $\erf z\equiv \frac{2}{\sqrt{\pi}}\int_0^z e^{-s^2}ds$ is the familiar error function \cite{abramowitz1972book}. For non-ideal qubits with a finite $T_1$, the modified single-shot readout fidelity at time $t\ll T_1$ is given by: 
\begin{equation}
\mathscr{F}(t) = \exp\left(-\frac{t}{2T_1}\right)\erf\left(\frac{\mbox{SNR}(t)}{\sqrt{2}}\right).
\end{equation}
It incorporates both the relaxation process of the qubit and the impact of noise on the readout. Notably, the single-shot readout fidelity depends only on the spin relaxation time $T_1$, as the dispersive Hamiltonian \eqref{Dispersive} involves only the population difference between the two qubit states.

\textit{The results}.--- In Fig.\:\ref{fig:S1} we show that employing squeezing can enhance SNR and reduce measurement time.  Specifically, Fig.\:2a reveals that using current technology ($\kappa = \chi_s=2\pi\times 0.15\:\mbox{MHz}$) and a displaced squeezed vacuum state ($\theta_\alpha=0$ and $\theta_\xi=\pi$) as input, there is an apparent SNR enhancement in the sub-microsecond temporal regime. For example, there is a nearly 50 percent enhancement in the SNR at around $0.9\:\mu\mbox{s}$, achieved by employing a displaced squeezed vacuum state with around 30 photons and a moderate squeezing parameter of $r=0.85$ (equivalent to 7.38 dB in decibels). Fig.\:2b shows that the chosen displaced squeezed vacuum state yields a notable readout fidelity of 97\% at around  $1\:\mu\mbox{s}$. Fig.\:2c indicates that by increasing the leakage rate slightly to $\kappa=2\chi_s$, the same SNR can be achieved within a shorter measurement time. Consequently, as depicted in Fig.\:2d, a notable readout fidelity of 97\% is attained at around $0.8\:\mu\mbox{s}$. 

Moreover, Fig.\:\ref{fig:S4} reveals that a phase mismatch between the local oscillator and the squeezed state consistently lowers both the SNR and fidelity in the sub-microsecond time regimes. At a fixed measurement time $t\approx 0.714\:\mu\mbox{s}$, and for a constant phase mismatch $\Delta\theta\equiv \varphi-\theta_\xi/2=0$, the SNR peaks when $r\approx 0.74$ (equivalent to 6.43 dB in decibels). It also shows that for a fixed $r\approx 0.74$, the SNR is enlarged from 3 to 3.5 by fixing $\Delta\theta$ as a multiple of $\pi$. It is important to note that when the phase matching condition $\theta_\xi-2\theta_\alpha=\pm \pi$ is fulfilled, the same SNR and fidelity can also be achieved using a squeezed coherent state $|\xi,\gamma\rangle$ with a displacement amplitude $\gamma=\alpha e^{-r}$. Consequently, as illustrated in Fig.\:3c, the readout fidelity peaks when the phase mismatch $\Delta\theta$ is a multiple of $\pi$. When the phase matching condition is fulfilled, Fig.\:3d shows that the fidelity peaks when $r\approx 0.74$, and declines when the degree of squeezing is either too high or too low.

The results presented in Fig.\:\ref{fig:S4} challenge the seemingly reasonable assumption that measurement accuracy can continue to improve as the degree of squeezing increases. In a quantum measurement with a single-mode photonic state with a small number of photons, the contributions from the anti-squeezing quadrature to the signal and noise are unavoidable. These contributions lead to the eventual decline of SNR and measurement fidelity as squeezing parameter $r$ increases. While the utilization of two or more photon modes theoretically allows for a higher signal-to-noise ratio by selecting two commuting quadratures, it necessitates intricate circuit architecture demands, such as incorporating two modes with opposing dispersive coupling constants \cite{didier2015heisenberg, govia2017enhanced}.

The required squeezing in the current proposal is well within reach by the current state-of-the-art experiments, where squeezing factor ranges from a few decibels to several dozens of decibels have been achieved, depending on different squeezing mechanisms. For instance, a squeezed state of microwave radiation with a squeezing factor of up to 8 dB has been reported using mechanical oscillator \cite{ockeloen2017noiseless}, while a broadband squeezed microwave radiation with a squeezing factor of up to 11.35 dB for a single-mode field has been reported using a Josephson traveling-wave parametric amplifier \cite{qiu2022broadband}.

\textit{Back-action on the qubit}.--- In the above discussion, we assume that the qubit properties such as its relaxation time is not affected by the probing photons. We now analyze how these photons, especially their squeezing, may affect qubit relaxation.

When the noise correlation time is much shorter than the spin decay time, the Bloch equation enables us to evaluate the back-action of the microwave on the spin qubit. The Hamiltonian \eqref{Dispersive} can be written as $H_s = \frac{1}{2}B_z\sigma_z$, where $B_z\equiv -\chi_s(a^\dagger a +aa^\dagger)$ is the effective longitudinal magnetic field in the rotating frame. In such a case, the spin qubit exhibits an infinite $T_1$ relaxation time, while its $T_2$ (transverse) relaxation time is determined by the noise correlation: $T_2^{-1}\equiv \int_0^t\langle \delta B_z(s)\delta B_z(t)\rangle ds$. Here, $\delta B_z\equiv B_z-\langle B_z\rangle$ is the effective magnetic noise experienced by the spin qubit. For a moderate displacement amplitude, the inclusion of squeezing results in a reduction in the spin $T_2$ relaxation time by a factor of $e^{2r}$.

As evident from the discussion above, the detuning responsible for generating dispersive coupling also contributes to suppressing the back-action. For an accurate evaluation of the spin relaxation time, it is imperative to revisit the original spin-photon coupling Hamiltonian, which describes the direct energy exchange between the spin qubit and the resonator via the absorption/emission of a resonator photon. Up to the leading order in spin-photon coupling $g_s$, the effective driving Hamiltonian takes the form \cite{d2019optimal}: $i\sqrt{\kappa}g_s\Delta^{-1}(b_{\mathrm{in}}^+\sigma_--b_{\mathrm{in}}\sigma_+)$, where $\Delta$ is the spin-resonator detuning. Consequently, the modified Hamiltonian for the spin qubit is $H_s=\frac{1}{2}\mathbf{B}\cdot\boldsymbol{\sigma}$, with $(B_x,B_y)\equiv \sqrt{2\kappa} g_s\Delta^{-1}(P_{\mathrm{in}},Q_{\mathrm{in}})$ as the effective transverse magnetic field in the rotating frame. The spin relaxation rate due to the coupling to the resonator is determined by the effective transverse magnetic field noise correlation via $T_1^{-1}\equiv \int_0^t\langle \delta B_x(s)\delta B_x(t)+\delta B_y(s)\delta B_y(t)\rangle ds$. A straightforward computation yields (See Supplemental Material)
\begin{equation}\label{T1} 
\frac{1}{T_1}= 2\gamma_{pu}\cosh 2r,
\end{equation}
where $\gamma_{pu} \equiv \kappa g_s^2\Delta^{-2}$ is the Purcell relaxation rate, characterizing the emission of a resonator photon into the environment. Eq.\:\eqref{T1} shows that the qubit relaxation rate is proportional to the Purcell relaxation rate, as expected, but is modified by the presence of squeezing. Specifically, the relaxation rate is a monotonically increasing function of the degree of squeezing, regardless of the displacement amplitude $|\alpha|$ when $|\alpha|$ is not too large. For $r=1$, one obtains $T_1^{-1}\approx 7.52 \gamma_{pu}$. Notice that for a strong probe, there are corrections to the total relaxation rate due to spin transitions induced by probe photons, but these can be neglected when the photon number is much smaller than a critical photon number $n_c = \Delta^2/4g_s^2$ \cite{d2019optimal}. 

\textit{Conclusion}.--- In this study, we demonstrate the effectiveness of employing displaced squeezed vacuum states to enhance the readout fidelity and reduce the readout time of a single spin in a semiconductor quantum dot through dispersive coupling to a resonator. In general, probing photons in a displaced squeezed vacuum state does not inherently contribute to the improvement of signal to noise ratio in spin measurement. The critical factor lies in a set of phase matching conditions among the squeezing phase, the coherent displacement phase, and the local oscillator phase during homodyne detection. When these phase matching conditions are met, a moderate degree of squeezing (a few decibels) proves effective in reducing the spin readout time to the sub-microsecond range while maintaining high fidelity. For instance, with current technology ($\kappa =\chi_s\approx  2\pi \times 0.15 \mbox{MHz}$), using a displaced squeezed vacuum state with $30$ photons and a mild squeezing parameter ($r = 0.85$, approximately 7.38 dB) allows us to achieve a readout fidelity of $97\%$ within a readout time of about $1\:\mu\mbox{s}$. Under the same condition, with a slightly larger leakage rate for the resonator ($\kappa =2\chi_s$), the same readout fidelity can be attained in approximately $0.8\:\mu\mbox{s}$.  Interestingly, we also find that squeezing beyond a certain threshold starts to hinder rather than help spin measurement, due to the inevitable contribution from the anti-squeezing quadrature to both the signal and noise.

\begin{acknowledgements}
We acknowledge financial support by US ARO via grants W911NF1710257 and W911NF2310018. We gratefully acknowledge the valuable discussion with Bill Coish.
\end{acknowledgements}

\end{document}